\pdfoutput=1
\documentclass[%
 reprint,
superscriptaddress,
nofootinbib,
 amsmath,amssymb,
 aps,
pra,
]{revtex4-1}
\usepackage{graphicx}
\usepackage{bm}
\usepackage{hyperref}
\usepackage[mathlines]{lineno}
\usepackage[utf8]{inputenc}
\usepackage{amsmath,amssymb}
\usepackage[T1]{fontenc} 
\usepackage{microtype} 
\usepackage[english]{babel} 

\usepackage{booktabs} 

\usepackage{graphicx}
\usepackage{amssymb}
\usepackage{braket,mleftright}
\usepackage{empheq}
\usepackage{subfigure}
\usepackage{stackrel}
\usepackage{soul}
\usepackage{blkarray}
\usepackage{multirow}
\usepackage{amsmath}
\usepackage{physics}
\usepackage{amsfonts}
\usepackage{bm}
\usepackage{bbold} 
\usepackage{color}
\usepackage{natbib}
\setcitestyle{square, comma, numbers,sort&compress, super}

\newcommand{\beq}{\begin{equation}}
\newcommand{\eeq}{\end{equation}}
\newcommand{\bse}{\begin{subequations}}
\newcommand{\ese}{\end{subequations}}
\newcommand{\bea}{\begin{eqnarray}}
\newcommand{\eea}{\end{eqnarray}}

\usepackage[utf8]{inputenc} 
\usepackage{amsmath}
\usepackage{enumitem} 
\setlist[itemize]{noitemsep} 

\usepackage{hyperref}
\usepackage{cleveref}
\usepackage{braket}
\usepackage{verbatim} 
\begin{document}


\title{Quantum metrology in a non-Markovian quantum evolution}

\author{Nicol\'{a}s Mirkin}
\email[Corresponding author:]{\,mirkin@df.uba.ar}
\affiliation{%
Departamento de F\'{i}sica “J. J. Giambiagi” and IFIBA, FCEyN, Universidad de Buenos Aires, 1428 Buenos Aires, Argentina
}%
\author{Martín Larocca}
\affiliation{%
Departamento de F\'{i}sica “J. J. Giambiagi” and IFIBA, FCEyN, Universidad de Buenos Aires, 1428 Buenos Aires, Argentina
}%

\author{Diego Wisniacki}

\affiliation{%
Departamento de F\'{i}sica “J. J. Giambiagi” and IFIBA, FCEyN, Universidad de Buenos Aires, 1428 Buenos Aires, Argentina
}%

\date{\today}%

\begin{abstract}
The problem of quantum metrology under the context of a particular non-Markovian quantum evolution is explored. 
We study the dynamics of the quantum Fisher information (QFI) of a composite quantum probe coupled to a Lorentzian environment, for a full variety of different classes of parameters.
We are able to find the best metrological state which is not maximally entangled but is the one which evolves fastest. This is shown by evidencing a connection between QFI and different quantum speed limits. At the same time, by optimizing a control field acting over the probes, we show how the total information flow is actively manipulated by the control such as to enhance the parameter estimation at a given final evolution time. At last, under this controlled scenario, a sharp interplay between the dynamics of QFI, non-Markovianity, and entanglement is revealed within different control schemes. 

\end{abstract}

\maketitle

\section{\label{Section-Intro}Introduction}
One of the most developed areas within the advent of quantum information technologies during the last two decades of quantum revolution has been undoubtedly quantum metrology. This field is responsible for the development of high-resolution and highly sensitive measurements of physical parameters, which is a central task for the prosperous evolution of technology \cite{bib:giovannetti2011advances,bib:degen2017quantum}. In this context, the question of whether the powerful resources of nature that have been revealed by quantum mechanics can be exploited to improve the precision in the estimation of unknown parameters associated with a quantum system has been deeply studied in the literature \cite{bib:bollinger1996optimal,bib:huelga1997improvement,bib:giovannetti2004quantum,bib:giovannetti2006quantum,bib:esteve2008squeezing,bib:paris2009quantum}.

Since any realistic quantum system interacts and exchanges information with an environment, the main challenge resides in tackling the problem of quantum metrology within the presence of decoherence and non-Markovianity (NM) \cite{bib:dorner2009optimal,bib:watanabe2010optimal,bib:lu2010quantum,bib:ma2011quantum,bib:chin2012quantum,bib:zhong2013fisher,bib:alipour2014quantum,bib:li2015classical,bib:benedetti2018quantum,bib:jahromi2019multiparameter,bib:sehdaran2019effect,bib:xu2019readout,bib:giorgi2019quantum,bib:gebbia2019two}. Owing both phenomena are related to the loss and gain of information, respectively, the question of how the dynamics of estimation is affected both by the presence of decoherence and NM is of paramount interest and worthy to study. In that sense, as memory effects -usually associated with non-Markovian quantum processes- allow to recover information from the environment that otherwise will be lost \cite{bib:medida1,bib:medida2}, this has opened a new door for applications in quantum metrology. 

A natural question arises then: is it possible to exploit some particular feature of the environment in order to enhance the precision of estimation of different classes of parameters? For instance, it has recently been shown that NM can be actively manipulated to generate a controlled degree of entanglement between two non-interacting subsystems coupled to the same non-Markovian reservoir \cite{bib:mirkin2019information,bib:mirkin_ent}; also that dissipation can be engineered to be a fully fledged resource for universal quantum computation \cite{bib:env_resource1}. So the possibility of engineering some particular feature of the environment to improve the process of estimating different classes of parameters seems certainly plausible \cite{bib:li2020non}.

In this work, we analyze a quantum metrology scenario within a particular non-Markovian quantum evolution. By using two non-interacting subsystems, we focus on the capability of this composite quantum system to act as a probe and to extract relevant information of parameters characterizing a common structured environment to which they are coupled as well as parameters characterizing the interaction. Seeking universality, we also focus on the process of estimation of the quantum probe itself. By exploring the role of initial entanglement within the composite probe in all the cases just mentioned, we find that the best metrological states are not the maximally entangled but the ones that evolve fastest. This is shown using the quantum speed limit (QSL), a tool that characterizes the minimum time a quantum system needs in order to travel a predetermined distance on Hilbert space \cite{bib:deffner2013quantum,bib:del2013quantum,bib:taddei2013quantum,bib:intro_speedup3,bib:mirkin2018comment}. Thus, while evidencing that the speed of evolution and the accuracy of the estimation are deeply connected under all the metrological situations covered, we also show that entanglement is not decisive to accelerate a non-Markovian evolution neither truly useful for improving the estimation of the parameter of interest.

Another important key-point of our work resides on the implementation of optimal control tools such as to achieve a controlled degree of precision on the estimation of the unknown parameter at a given final evolution time. Therefore, after identifying the best metrological states, we show how by optimizing a control field over the composite probe, the total information flowing throughout the evolution can be actively accommodated by the control such as to maximize the precision of the estimation at this given final evolution time. Finally, by exploring different control schemes, we reveal a direct dynamical relation between the information flows regarding the precision of estimation, NM and entanglement within some particular circumstances.   

The manuscript is organized as follows. First, Section II reviews the main concepts of quantum metrology theory. Then, Section III summarizes the concept of QSL for non-unitary dynamics by presenting two of the most well-known bounds that have been previously derived in the literature. Next, Section IV analyzes the main features characterizing NM together with a measure able to quantify it. Afterward, Section V provides the physical model in which we based our study and in Section VI the main results obtained for this model are presented. Lastly, Section VII concludes with some final remarks.

\section{Quantum Metrology Theory}
In this Section, we provide a brief summary of the most relevant features of quantum metrology theory. The purpose of quantum metrology is to deal with estimation processes within quantum systems, pursuing the best precision that is physically allowed \cite{bib:giovannetti2011advances,bib:degen2017quantum}. For instance, let us consider a situation in which the quantum evolution of a certain system is known unless for a certain parameter $\tilde{\lambda}$. This $\tilde{\lambda}$ may be estimated from the knowledge of the initial and final states of a given probe that undergoes the process of interest. The metrological procedure is usually the following: a quantum probe is first initialized in a particular input state and as it evolves is transformed into a mixed state encoding information of the unknown parameter $\tilde{\lambda}$. After the evolution, a suitable measurement must be done over the probe such as to extract information about it. Finally, each experimental result should be associated with some estimator of the parameter of interest. Under this context, any measurement of a certain observable X is associated with an outcome x which occurs with a conditional probability distribution $p_{X}(x|\tilde{\lambda})$, which is defined by \cite{bib:escher2011general,bib:salvatori2014quantum, bib:benedetti2018quantum}
\begin{equation}
    p_{X}(x|\tilde{\lambda})=\Tr\left(P_{x} \rho_{\tilde{\lambda}}\right).
\end{equation}
Above, $ \rho_{\tilde{\lambda}}$ refers to the quantum state of the probe and $P_{x}$ to positive operator-valued measures, satisfying the relation $\sum_{x}P_{x}=\mathbb{1}$, usually known in the literature as a POVM. Therefore, in order to estimate the value of our unknown parameter $\tilde{\lambda}$ from the outcome measurements, an estimator is needed. This estimator must be a function of the measurement outcomes, i.e. $\hat{\tilde{\lambda}}=\hat{\tilde{\lambda}}\left(x_1,x_2,... \right)$ and also should satisfy certain properties, such as being unbiased 
\begin{equation}
    E \left[ \hat{\tilde{\lambda}}-\tilde{\lambda} \right]= \prod_{i}\sum_{x_{i}}\hat{\tilde{\lambda}}\left(x_1,...,x_n \right)-\tilde{\lambda}=0 \quad \forall \tilde{\lambda},
\end{equation}
where $E[...]$ corresponds to the mean with respect to the $n$ outcomes $x_i$ and $\tilde{\lambda}$ denotes the true value of the parameter. Moreover, it is also important to require a small variance for the estimator, i.e. $\mathrm{Var}\left(\tilde{\lambda},\hat{\tilde{\lambda}}\right)=E[\hat{\tilde{\lambda}}^{2}]-E[\tilde{\lambda}]^{2}$, considering this quantity measures the overall precision of the inference process \cite{bib:salvatori2014quantum}. In regards to this quantity, it is well known that a lower bound for the variance of any estimator is imposed by the Cramér-Rao theorem \cite{bib:helstrom1976quantum,bib:holevo2011probabilistic}, 
\begin{equation}
    \mathrm{Var}\left(\tilde{\lambda},\hat{\tilde{\lambda}}\right) \geq \frac{1}{M G_{\tilde{\lambda}}},
\end{equation}
where M denotes the number of independent measurements and $G_{\tilde{\lambda}}$ is known as the Fisher information (FI) and is defined by
\begin{equation}
    G_{\tilde{\lambda}}=\sum_{x}\frac{[\partial_{\tilde{\lambda}}\,p_{X}(x|\tilde{\lambda}) ]^{2}}{p_{X}(x|\tilde{\lambda})}.
\label{povm}
\end{equation}

The challenge is then to choose the best estimator such as to achieve an optimal inference and saturate the Cramér-Rao bound. The fact that different observables will lead to different probability distributions is intuitive, which means that each one will be associated with a particular FI and hence to different precisions for the estimation of the unknown parameter $\tilde{\lambda}$ \cite{bib:paris2009quantum}. The ultimate bound is traditionally obtained upon maximizing the FI over all the set of possible POVM's. The best measurement that provides the maximum precision is quantified with what is called the quantum Fisher information (QFI), which is given by

\begin{equation}
    \mathcal{F}_{\tilde{\lambda}}=\sum_{n}\frac{\left(\partial_{\tilde{\lambda}} \omega_n \right)^{2}}{\omega_{n}}+2\sum_{n\neq m}\frac{(\omega_n-\omega_m)^{2}}{\omega_n+\omega_m}||\bra{\psi_n}{\partial_{\tilde{\lambda}} \psi_m \rangle}||^{2},
\end{equation}
where $\{ \omega_{n} \}$ are the eigenvalues of the reduced density matrix of the probe and $\{ \ket{\psi_{n}} \}$ its eigenvectors. In this way, the QFI is lower bounded by the FI, i.e. $G_{\tilde{\lambda}}\leq \mathcal{F}_{\tilde{\lambda}}$. Note that $\tilde{\lambda}$ could be any parameter characterizing either the probe, the interaction or even the environment. Let us finally stress that since $\mathcal{F}_{\tilde{\lambda}}(t)$ is truly a dynamical quantity, in this work we will sometimes work with the total QFI throughout a particular given evolution, i.e.
\begin{equation}
\mathcal{F}_{\tilde{\lambda}}^{(tot)}=\int_{0}^{T}\mathcal{F}_{\tilde{\lambda}}(t')dt',
\label{f_tot}
\end{equation}
with $T$ referring to some fixed evolution time. In general, $\mathcal{F}_{\tilde{\lambda}}^{(tot)}$ has no meaning in quantum metrology, since what is really important is the QFI at a fixed evolution time $\mathcal{F}_{\tilde{\lambda}}(T)$. However, by using optimal control tools, we will show that these two quantities are strongly related. 

\section{Quantum Speed Limits for open quantum evolutions}
In this Section, we review two of the most well-known QSL's previously derived in the literature for non-unitary quantum evolutions. The QSL time $\tau$ is defined as the minimal time a quantum system needs in order to evolve from an initial to a final state, separated by a given predetermined distance \cite{bib:deffner2013quantum,bib:del2013quantum,bib:taddei2013quantum,bib:intro_speedup3,bib:mirkin2018comment}. 
The first approach we present towards the correct formulation of the QSL, is based on the definition of the Bures fidelity between an initial and a final state, i.e. \cite{bib:taddei2013quantum,bib:intro_speedup3}
\begin{equation}
    F_{B}(\rho_0,\rho_t)=\Tr \left(\sqrt{\sqrt{\rho_0}\rho_t\sqrt{\rho_0}} \right).
\end{equation}
It can be proven that the tightest lower bound for the actual path length of the evolution is given by the Bures angle
\begin{equation}
    \mathcal{L}(\rho_0, \rho_t) \equiv \arccos{(F_{B}(\rho_0,\rho_t))} \leq \int_{0}^{t}\sqrt{\frac{\mathcal{F}_{t}(t')}{4}}dt',
\end{equation}
where $\mathcal{F}_{t}(t)$ corresponds to the QFI for time estimation and $\mathcal{L}(\rho_0, \rho_t)$ to the Bures angle, which is a predetermined distance (i.e. between orthogonal states $\mathcal{L}(\rho_0, \rho_t)=\pi/2$). Therefore, since $\sqrt{\frac{\mathcal{F}_{t}(t)}{4}}$ is commonly regarded as the instantaneous speed of evolution \cite{bib:taddei2013quantum}, the time that saturates that fixed predetermined distance defines the minimum time of evolution, the QSL, that we name $\tau_{\mathcal{F}}$,
\begin{equation}
    \mathcal{L}(\rho_0, \rho_t) = \int_{0}^{\tau_{\mathcal{F}}}\sqrt{\frac{\mathcal{F}_{t}(t')}{4}}dt'. 
\label{taddei}
\end{equation}
In other words, $\tau_{\mathcal{F}}$ reflects the time that the system takes to travel -along the actual evolution path- the same length as the geodesic’s length between two different predetermined states. Moreover, as it has already been proven in Ref. \cite{bib:intro_speedup3}, this expression for the QSL is the only one that sticks close to the essence of the QSL theory \cite{bib:mandelstam1945energy,bib:margolus1998maximum,bib:levitin2009fundamental} since it is always possible to find an evolutionary path that, for every time, saturates the bound. This will occur whenever the system evolution equals the geodesic path.

Another very popular approach used in the literature to derive an expression for the QSL is the one proposed by Deffner and Lutz \cite{bib:deffner2013quantum}, based on the von Neumann trace inequality for Hilbert-Schmidt class operators. The tightest QSL they found can be consistently defined as
\begin{equation}
    \sin^{2}\left(\mathcal{L}(\rho_0, \rho_t)\right)=\int_{0}^{\tau_{op}}||\dot{\rho}(t')||_{op} \, dt',
\label{deffner}
\end{equation}
where $||A||_{op}$ is the operator norm of A. Similarly, the time that saturates the distance fixed by the l.h.s. of Eq. (\ref{deffner}) corresponds to the QSL time and we note it $\tau_{op}$. However, as it has also been demonstrated in Ref. \cite{bib:intro_speedup3}, under this approach it turns impossible to find an evolutionary path where Eq. (\ref{deffner}) is saturated at all times.

Let us remark that both expressions presented above for the QSL will be used in Section VI to illustrate how the speed of evolution and the accuracy of the estimation are closely related within our non-Markovian quantum metrology scenario.

\section{Non-Markovianity measure}
There are many different ways to quantify NM, being one of the most popular approaches related to the revivals of distinguishability and originally proposed by Breuer, Laine, and Piilo (BLP) \cite{bib:medida1}. The distinguishability can be quantified by the derivative of the trace distance, which is defined as $\mathcal{D}(\rho_{1},\rho_{2})=\dfrac{1}{2}||\rho_{1}-\rho_{2}||$ and where $||A||=tr(\sqrt{A^{\dagger}A})$. 
Under a Markovian regime, quantum states become less and less distinguishable, there is a continuous loss of information to the environment. But on a non-Markovian regime, distinguishability between states can increase and this is equivalent to say that information is flowing from the environment back to the system. Therefore, BLP states that a quantum map is non-Markovian if there exists at least a pair of initial states $\rho_{1}(0)$ and $\rho_{2}(0)$ such that the distinguishability between them increases during some interval of time, i.e.
\begin{equation}
    \sigma(\rho_{1}(0),\rho_{2}(0),t)=\dfrac{d}{dt}\mathcal{D}(\rho_{1}(t),\rho_{2}(t)) > 0.
\label{distinguibilidad}
\end{equation}
This idea can also be extended to define a measure of the degree of NM in a quantum process via 
\begin{equation}
    \mathcal{N}_{BLP}=\max\limits_{{\lbrace\rho_{1}(0),\rho_{2}(0)\rbrace}} \int_{0, \sigma >0}^{T} \sigma \left (\rho_{1}(0),\rho_{2}(0),t'\right)  dt',
    \label{BLP}
\end{equation}
where T refers to the final evolution time of the process under consideration. 
In general, Eq. (\ref{BLP}) is integrated to infinity, but since here we will consider control protocols with a certain finite duration, we will quantify NM for a restricted time interval.

\section{\label{Section-Systems}Physical model}
The system we use as a platform to study different aspects of quantum metrology within a non-Markovian quantum evolution consists on two non-interacting two-level atoms, acting as a composite probe, coupled to a common zero-temperature bosonic reservoir composed by a set of $M$-harmonic oscillators \cite{bib:manis2,bib:lofranco4,bib:mirkin2019information}. The total microscopic Hamiltonian describing the model is given by
\begin{equation}
\begin{split}
    H & =  \ H_{S} + H_{E} + H_{int} \\  & = \sum_{i=1}^{2}\omega_{i}(t)\sigma_{+}^{(i)}\sigma_{-}^{(i)} + \sum_{k=1}^{M}\nu_{k}b^{\dagger}_{k}b_{k} 
    \\ & + \sum_{i=1}^{2}\left(a_{i}\sigma_{+}^{(i)}\otimes \sum_{k=1}^{M}g_{k}b_{k}+H.c \right),
\end{split}
\label{eq_hamilt}
\end{equation}
where $\sigma_{j}^{(i)}$ ($j=x,y,z$) correspond to the Pauli matrices of each atom ($i=1,2$), $\sigma_{\pm}^{(i)}=\dfrac{1}{2}(\sigma_{x}^{(i)}\pm i \sigma_{y}^{(i)})$, $b_{k}^{\dagger}$ and $b_{k}$ to the creation and annihilation operators, $g_{k}$ is the coupling constant to the k-th mode of the bath and $\nu_{k}$ its frequency, $a_{i}$ is a dimensionless coupling constant measuring the interaction with the reservoir, and finally $\omega_{i}(t)$ refers to the energy difference between the ground $\ket{0}$ and excited state $\ket{1}$ of the atom $\textit{i}$, which we assume to be time dependent and of the form
\begin{equation}
    \omega_{i}(t)=\omega_{0}+\epsilon_{i}(t).
\end{equation}
In principle, $\epsilon_{i}(t)$ is an arbitrary driving field over the atom $\textit{i}$ but for a matter of simplicity we will work under the framework of Global Addressing, where $\epsilon_{1}(t)=\epsilon_{2}(t)=\epsilon(t)$. Assuming that initially the environment has no excitations ($\ket{0_{B}}$) and that the dynamics is restricted to one excitation in the k-th mode, the whole initial state of the total system is
\begin{equation}
\ket{\psi(0)}=\left(C_{01}\ket{10}+C_{02}\ket{01} \right)\otimes_{k}\ket{0_{B}},
\end{equation}
and therefore its dynamics is given by
\begin{equation}
\begin{split}
\ket{\psi(t)} &=\ C_1(t)\ket{10}\ket{0_{B}}+C_2(t)\ket{01}\ket{0_{B}} \\ & +\sum_{k}C_{k}(t)\ket{00}\ket{k_{B}},
\end{split}
\end{equation} 
being $\ket{k_{B}}$ the state of the reservoir with only one excitation in the k-th mode ($\ket{k_{B}}=b^{\dagger}_{k}\ket{0_{B}}$). The next step is to take the continuum limit for the environment, by assuming a Lorentzian spectral density of the form
\begin{equation}
J(\nu)=\frac{\mathcal{R}^{2}}{\pi a_{t}^{2}}\frac{\lambda}{(\nu-\omega_0)^{2}+\lambda^{2}},    
\end{equation}
where $\mathcal{R}$ is the vacuum Rabi frequency, $a_t$ an effective coupling constant defined as $a_t=\sqrt{a_{1}^{2}+a_{2}^{2}}$ and $\lambda$ the width of the spectral density of the bath. Finally, is straightforward to follow the procedure put forward in Ref. \cite{bib:mirkin2019information} and derive these two coupled differential Eqs. for $C_1(t)$ and $C_2(t)$, respectively 
\begin{equation}
    \Ddot{C}_{1}+(\lambda-i\epsilon(t))\dot{C}_{1}+\left(\frac{a_{1}\mathcal{R}}{a_t}\right)^{2}C_{1}+a_{1}a_{2}\left(\frac{\mathcal{R}}{a_t}\right)^{2}C_{2}=0
\label{eqc1}
\end{equation}
\begin{equation}
    \Ddot{C}_{2}+(\lambda-i\epsilon(t))\dot{C}_{2}+\left(\frac{a_{2}\mathcal{R}}{a_t}\right)^{2}C_{2}+a_{1}a_{2}\left(\frac{\mathcal{R}}{a_t}\right)^{2}C_{1}=0.
\label{eqc2}
\end{equation}
The density matrix can be written as \cite{bib:manis2, bib:lofranco4}
\begin{equation}
    \rho(t)=\begin{pmatrix}
0 & 0 & 0 & 0 \\
0 & \abs{C_{1}(t)}^{2} & C_{1}(t)C^{*}_{2}(t) & 0 \\
0 & C^{*}_{1}(t)C_{2}(t) & \abs{C_{2}(t)}^{2} & 0 \\
0 & 0 & 0 & 1-\abs{C_{1}(t)}^{2}-\abs{C_{2}(t)}^{2}
\end{pmatrix}.  
\end{equation}

Before proceeding with the results, we shall stop to accentuate two important details. First, it is important to stress that since we are considering initial states of the form $\ket{\psi(0)}=C_{01}\ket{10}+C_{02}\ket{01}$, it is possible to parametrize the initial coefficients as $C_{01}=\sqrt{\frac{1-s}{2}}$ and $C_{02}=\sqrt{\frac{1+s}{2}}e^{i\phi}$, where $-1\leq s \leq 1$ and $0\leq \phi \leq \pi$. Note that in the case in which $s=0$, the initial state is entangled and if $|s|=1$ it is separable, so we will refer to parameter $s$ as the \textit{initial separability}. Secondly, it is critical to mention that there exist some very specific initial states. For instance, if $\ket{\psi(0)}=(a_2/a_t)\ket{10}-(a_1/a_t)\ket{01}$, the state is named \textit{sub-radiant} and is a constant solution of Eqs. (\ref{eqc1}) and (\ref{eqc2}) that does not decay in time. On the contrary, if $\ket{\psi(0)}=(a_1/a_t)\ket{10}+(a_2/a_t)\ket{01}$ the state is orthogonal to the previous one, is called \textit{super-radiant} and is the one that evolves fastest, as will be shown later.

\section{Results}
\subsection{Quantum metrology: QSL, entanglement and optimal control}

\begin{figure}[!htb]
\begin{center}
\includegraphics[width=88mm]{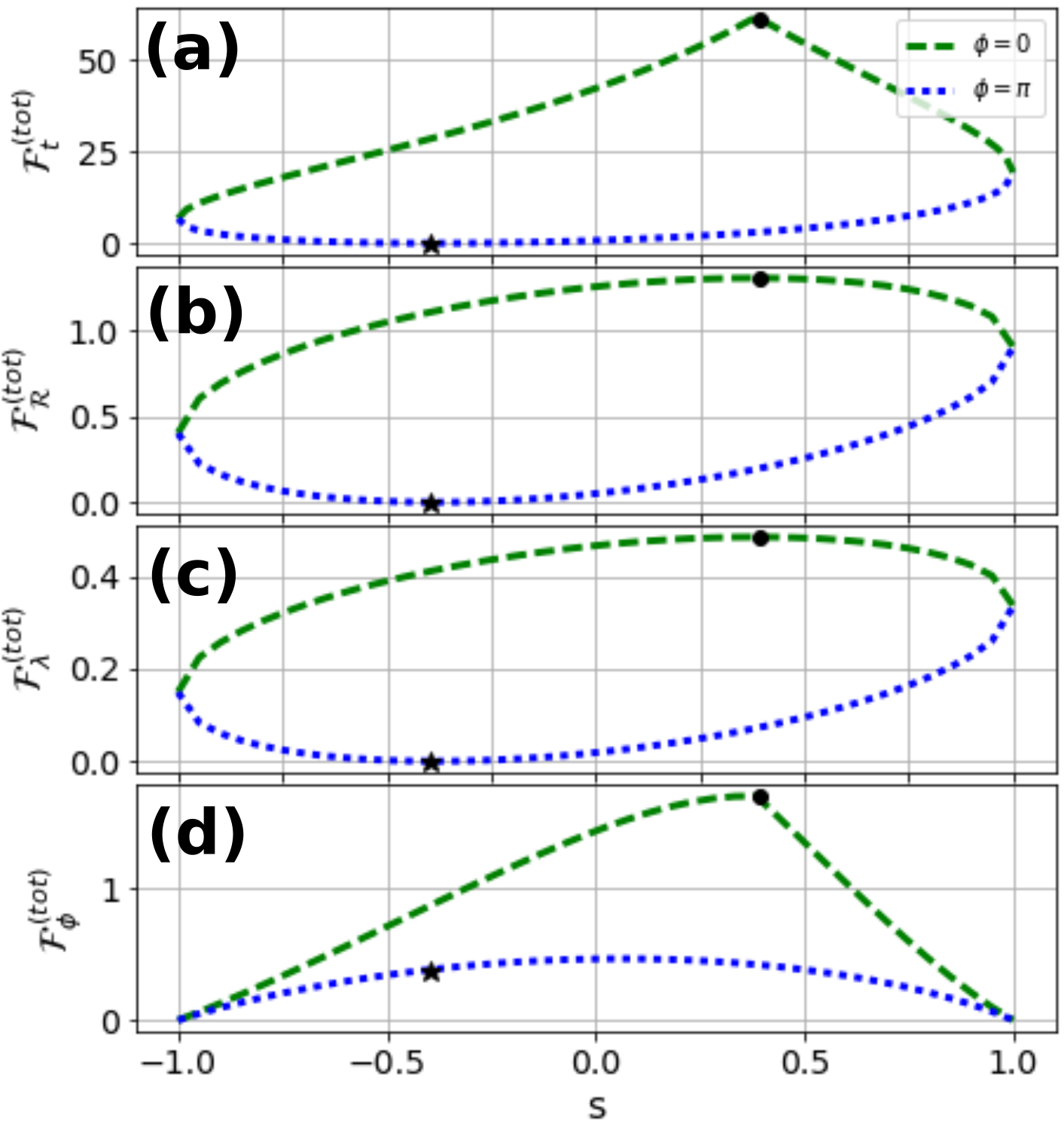}
\begin{footnotesize}
\caption{\textbf{(a)}: $\mathcal{F}_{t}^{(tot)}$ for time estimation \textbf{(b)}: $\mathcal{F}_{\mathcal{R}}^{(tot)}$ for estimating the interaction parameter $\mathcal{R}$ \textbf{(c)}: $\mathcal{F}_{\lambda}^{(tot)}$ for environment estimation \textbf{(d)}: $\mathcal{F}_{\phi}^{(tot)}$ for phase estimation. In all panels, the black asterisk corresponds to the \textit{sub-radiant} state ($s^{*}\eqsim -0.38$, $\phi=\pi$), while the black dot corresponds to the \textit{super-radiant} state ($s^{*}\eqsim 0.38$, $\phi=0$). Interaction parameters are fixed as $a_1=0.4$,  $a_2=0.6$, $\mathcal{R}=5$ and $T=2$, while $\lambda$ is set equal to 1. Any other initial state with a different $\phi$ will reside inside the topological structures found above.}
\label{met1}
\end{footnotesize}
\end{center}
\end{figure}

In the first part of this Section, we will identify the best metrological state to estimate the most general variety of parameters, i.e. characterizing either the environment, the interaction or even the quantum probe itself. Let us remark that, at this point, we will consider the best metrological state as the one that achieves the greatest $\mathcal{F}_{\tilde{\lambda}}^{(tot)}$ throughout an evolution\footnote{The total QFI gives us a sort of average of the information that could be obtained throughout the dynamics. More $\mathcal{F}^{(tot)}_{\tilde{\lambda}}$ implies a greater possibility of achieving a better degree of precision at some particular time. However, despite it is true that usually this quantity has no much physical meaning in quantum metrology, we will show with optimal control tools that this quantity is sharply related with the maximum QFI achievable by optimizing a control field for a fixed given evolution time.}. Under this framework, a natural question arises then: is there a common feature characterizing the best metrological state for estimating each different subset of parameters? For instance, is entanglement truly useful for improving the estimation or is there a more fundamental physical reason that can be exploited? With this question in mind, in Fig. \ref{met1} we study the $\mathcal{F}_{\tilde{\lambda}}^{(tot)}$ for different classes of parameters, either characterizing the interaction (time $t$ and vacuum Rabi frequency $\mathcal{R}$), the environment (width $\lambda$ of the Lorentzian spectral density) but also the quantum probe itself (initial phase $\phi$). All these quantities are plotted as a function of the initial separability s, fixing the interaction parameters as $a_1=0.4$, $a_2=0.6$, $\mathcal{R}=5$ and $T=2$. We point out that the same results were obtained by fixing other interaction parameters.


\renewcommand{\figurename}{Figure} 

As it is clear from Fig. \ref{met1}, the best initial metrological state is the \textit{super-radiant} state since it is the one that  maximizes the $\mathcal{F}_{\tilde{\lambda}}^{(tot)}$, either for estimating a parameter of the interaction (panels \textbf{(a)} and \textbf{(b)}), a parameter of the environment (panel \textbf{(c)}) or even a parameter describing the initial state of the quantum probe itself (panel \textbf{(d)}). On the other hand, as can be intuitively deduced from what was expressed in Section V, considering that the \textit{sub-radiant state} does not decay in time, it is not surprising that this state cannot extract information from the interaction or from the environment. This is the reason why $\mathcal{F}_{\tilde{\lambda}}^{(tot)}$ is zero for the situations covered by panels \textbf{(a)}, \textbf{(b)} and \textbf{(c)}. Nevertheless, when a parameter of the initial state of the probe is being estimated, such as the initial phase $\phi$, no interaction with the environment is needed and so $\mathcal{F}_{\phi}^{(tot)}$ is not zero for this particular initial state, as can be observed from the asterisk on the blue dotted curve of panel \textbf{(d)}.

\begin{figure}[!htb]
\begin{center}
\includegraphics[width=82mm]{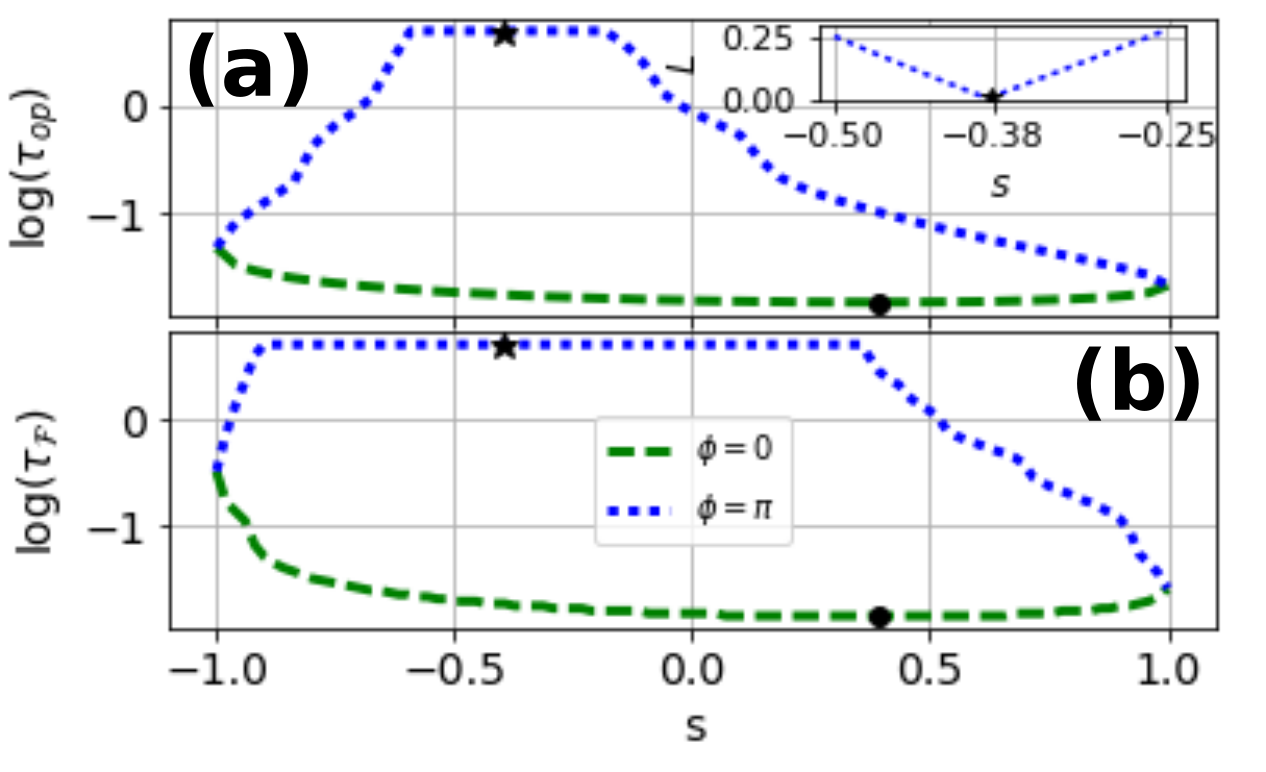}
\begin{footnotesize}
\caption{All quantities are plotted as a function of the initial separability s and for two different initial $\phi$. \textbf{(a)}: $\log{(\tau_{op})}$ given by Eq. (\ref{deffner}) and fixing $\mathcal{L}=\pi/4$. The inset corresponds to the actual distance $L$ (quantified by the r.h.s. of Eq. (\ref{deffner})) travelled by the system within the region near the \textit{sub-radiant} state, where the evolution is too slow to reach the predetermined distance given by the geodesic. \textbf{(b)}: $\log{(\tau_{\mathcal{F}})}$ given by Eq. (\ref{taddei}) and fixing $\mathcal{L}=\pi/4$. In both panels, the black asterisk corresponds to the \textit{sub-radiant} state ($s^{*}\eqsim -0.38$, $\phi=\pi$), while the black dot corresponds to the \textit{super-radiant} state ($s^{*}\eqsim 0.38$, $\phi=0$). Parameters are set the same as Fig. \ref{met1}. Same results were obtained by fixing other interaction parameters and Bures distance $\mathcal{L}$.}
\label{qsl1}
\end{footnotesize}
\end{center}
\end{figure}

Until now, we have identified the best initial metrological state for a whole set of different physical parameters and verified that this state is not the maximally entangled one. This is consistent with the fact that not all entangled quantum states are useful for quantum metrology and that often suffer more from certain non-unitary processes \cite{bib:hyllus2010not,bib:frowis2014optimal,bib:altenburg2016optimized}. Thus, the question of which is the physical reason underlying the best precision of estimation in our system still remains unanswered. As an approach to this problem, we shall return to what has already been pointed out in Section III, regarding that there is a relation between the QSL given by Eq. (\ref{taddei}) and the total QFI for time estimation $\mathcal{F}_{t}^{(tot)}$. This relation implies that the QFI for time estimation can be interpreted as a measure of the speed of evolution. Therefore, it is obvious that the initial state that maximizes $\mathcal{F}_{t}^{(tot)}$ will at the same time minimize the QSL defined by Eq. (\ref{taddei}). However, while this is true for the QFI for time estimation, this is certainly not obvious for the cases where other parameters are being estimated, such as $\mathcal{R}$, $\lambda$ or $\phi$. For example, while the speed of evolution of the \textit{sub-radiant} state is zero for all time, we still have a non-zero $\mathcal{F}_{\phi}^{(tot)}$ for phase estimation, as it can be observed from the blue dotted curve in panel \textbf{(d)} of Fig. 1. For this reason, since these other quantities are not related (in principle) with the speed of evolution, to explore thoroughly the interplay between the accuracy of estimation of different parameters and the speed of evolution seems something worthy to do. As a consequence, in Fig. \ref{qsl1} we plot $\tau_{\mathcal{F}}$ but also $\tau_{op}$, which is a QSL of different nature (i.e. not based on the QFI), for the same set of parameters of Fig. \ref{met1}. We remark that the same results were obtained by fixing other interaction parameters and Bures distance.    

\renewcommand{\figurename}{Figure} 

\begin{figure}[!htb]
\begin{center}
\includegraphics[width=80mm]{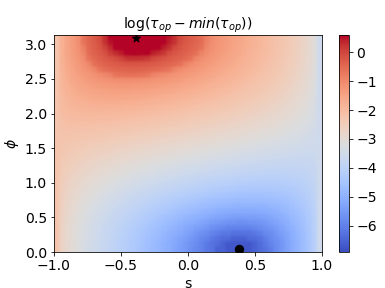}
\begin{footnotesize}
\caption{The $\tau_{op}$ in logarithmic scale, having previously subtracted the minimum value from all values so that the regions are well contrasted. The black asterisk corresponds to the \textit{sub-radiant} state while the black dot to the \textit{super-radiant} state. Parameters are set the same as Fig. \ref{met1}. Same results were obtained by fixing other interaction parameters and Bures distance $\mathcal{L}$.}
\label{qsl2}
\end{footnotesize}
\end{center}
\end{figure}

The first remarkable thing to notice from Fig. \ref{qsl1} is that the two QSL's have a similar behaviour, despite their different nature. The two quantities are able to identify both the $\textit{sub-radiant}$ and $\textit{super-radiant}$ states. While the \textit{sub-radiant} state takes infinite time to travel a given predetermined distance since it does not decay in time, the \textit{super-radiant} state is the one that evolves fastest. As can be seen, the QSL given by $\tau_{op}$ proves to be more sensitive to identify the slow states such as the ones near the \textit{sub-radiant}. In this region, states are so slow that although they are allowed to evolve during a long time, they cannot reach the predetermined distance established by the geodesic, as it is illustrated in the inset of panel \textbf{(a)}. For this reason, we can use the $\tau_{op}$ as a tool to explore with more detail the speed of evolution of all possible initial states and then identify more precisely the \textit{sub-radiant} state, as it is shown in Fig. \ref{qsl2}.

\renewcommand{\figurename}{Figure} 

As it is clear from all the above Figures, the best initial metrological state is the \textit{super-radiant} state, since it is not only the fastest but also which maximizes the $\mathcal{F}_{\tilde{\lambda}}^{(tot)}$, either for estimating a parameter characterizing the interaction, the environment or even the quantum probe itself. However, imagine you are an experimental physicist and you have to measure your observable at a given particular time $T$. Therefore, the fact that $\mathcal{F}_{\tilde{\lambda}}(t)$ has a dynamical behavior implies that if $\mathcal{F}_{\tilde{\lambda}}^{(tot)}$ is huge along the whole evolution this will be absolutely useless unless that, at that particular time T, $\mathcal{F}_{\tilde{\lambda}}(T)$ has a maximum. In this sense, a possible strategy could be to implement some sort of control field such as to maximize the final value $\mathcal{F}_{\tilde{\lambda}}(T)$ and consequently achieve a better degree of precision on the estimation of your parameter $\tilde{\lambda}$ at that particular final time in which the measurement is done. Under this context, an interesting question is: the fact that one has more $\mathcal{F}_{\tilde{\lambda}}^{(tot)}$ along a certain total evolution, necessarily implies that if one implements an optimization over the final value $\mathcal{F}_{\tilde{\lambda}}(T)$, this final value will be accordingly big? In other words, is the total information flow somehow accommodated by the control field? If this is true, then we should be able to reconstruct the same qualitative topology of some panels of Fig. \ref{met1}, by plotting the optimal final value $\mathcal{F}_{\tilde{\lambda}}(T)$ obtained by the optimization as a function of the initial separability s and for different $\phi$. 

\begin{figure}[!htb]
\begin{center}
\includegraphics[width=85mm,scale=0.99]{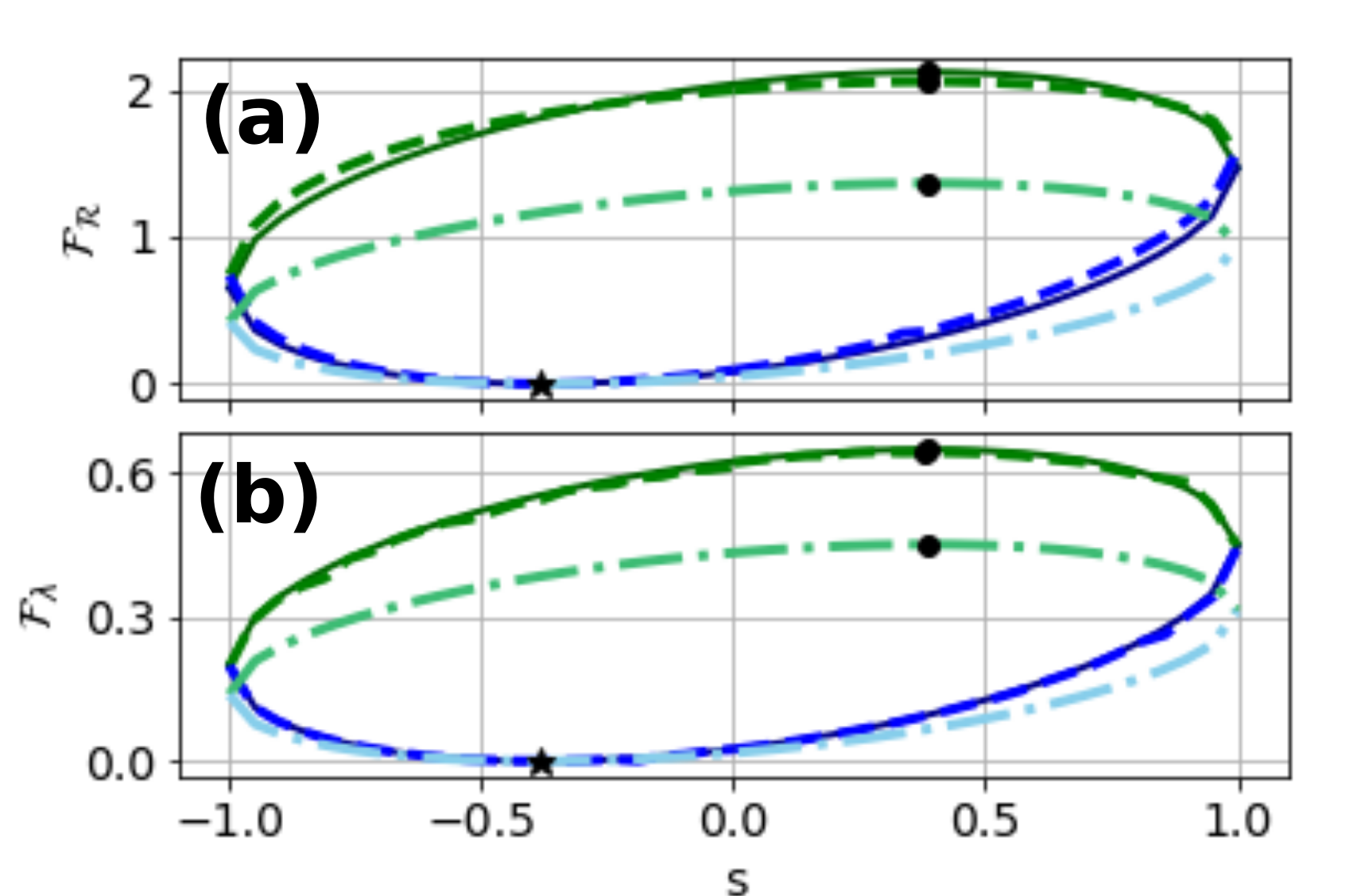}
\begin{footnotesize}
\caption{\textbf{(a)} QFI of the parameter $\mathcal{R}$ characterizing the interaction. \textbf{(b)} QFI of the width $\lambda$ of the spectral density of the environment. In both panels, all the green curves correspond to the case with $\phi=0$, while the blue ones are for $\phi=\pi$. At the same time, the dashed line refers to the optimal final value $\mathcal{F}_{\tilde{\lambda}}(T)$ obtained by the optimization; the dark solid line to the global maximum of $\mathcal{F}_{\tilde{\lambda}}(t)$ reached throughout the non-controlled evolution; and finally the light dotted-dashed line to $\mathcal{F}_{\tilde{\lambda}}^{(tot)}$ (given by Eq. (\ref{f_tot})), also throughout the non-controlled evolution. 
Interaction parameters are fixed as $a_1=0.4$,  $a_2=0.6$, $\mathcal{R}=10$ and $T=2$. Same results were obtained by fixing other interaction parameters.}
\label{optim_ov}
\end{footnotesize}
\end{center}
\end{figure}

In order to study this, we have numerically optimized the coupled differential Eqs. (\ref{eqc1}) and (\ref{eqc2}) to find an optimal field $\epsilon(t)$ that maximizes the functional $\mathcal{F}_{\tilde{\lambda}}(T)$. We have resorted to finite-length piece-wise constant controls, where the control function $\epsilon(t)$ was taken as a vector of control variables $\epsilon(t) \rightarrow \{ \epsilon_{k} \} \equiv \Vec{\epsilon}$, a field with constant amplitude $\epsilon_{k}$ for each time step \cite{bib:larocca2019exploiting}. The optimization was done by dividing the driving time T into 8 equidistant time steps  $(k=1,2,..., 8)$, exploring several random initial seeds and using standard optimization tools from the Python SCIPY library \cite{bib:scipy}. The optimal results obtained by numerical optimization, both to estimate the parameter $\mathcal{R}$ of the interaction as well as the width $\lambda$ of the Lorentzian spectral density of the environment for a given evolution time T are shown in Fig. \ref{optim_ov}.

\renewcommand{\figurename}{Figure} 

As can be observed from Fig. \ref{optim_ov} and was intuitively suggested before, the fact of having a certain degree of total QFI $\mathcal{F}_{\tilde{\lambda}}^{(tot)}$ throughout a whole process (see light dotted-dashed lines in Fig. \ref{optim_ov}), allows us to manipulate that total flow of information with a control field such as to have a maximum for $\mathcal{F}_{\tilde{\lambda}}(T)$ at some particular fixed evolution time $T$ which is of experimental interest (see dashed lines in Fig. \ref{optim_ov}). As can be noticed, the maximum value achievable with optimal control for $\mathcal{F}_{\tilde{\lambda}}(T)$ depends exclusively on the total degree of information that we had before within the non-controlled scenario. In addition, this maximum value achieved for $\mathcal{F}_{\tilde{\lambda}}(T)$ by optimizing the control field for a specific final time T coincides with the global maximum of $\mathcal{F}_{\tilde{\lambda}}(t)$ reached at an arbitrary time throughout the whole non-controlled evolution, as it is clear from comparing the dark solid lines with the dashed ones in Fig. \ref{optim_ov}. This is a numerical sample of the information flows being actively manipulated by the control field, a statement that will be explored with more detail in the next subsection.
 

\subsection{Exploiting information flows for quantum control}
\begin{figure*}[!htb]
\begin{center}
\includegraphics[width=150mm,scale=0.99]{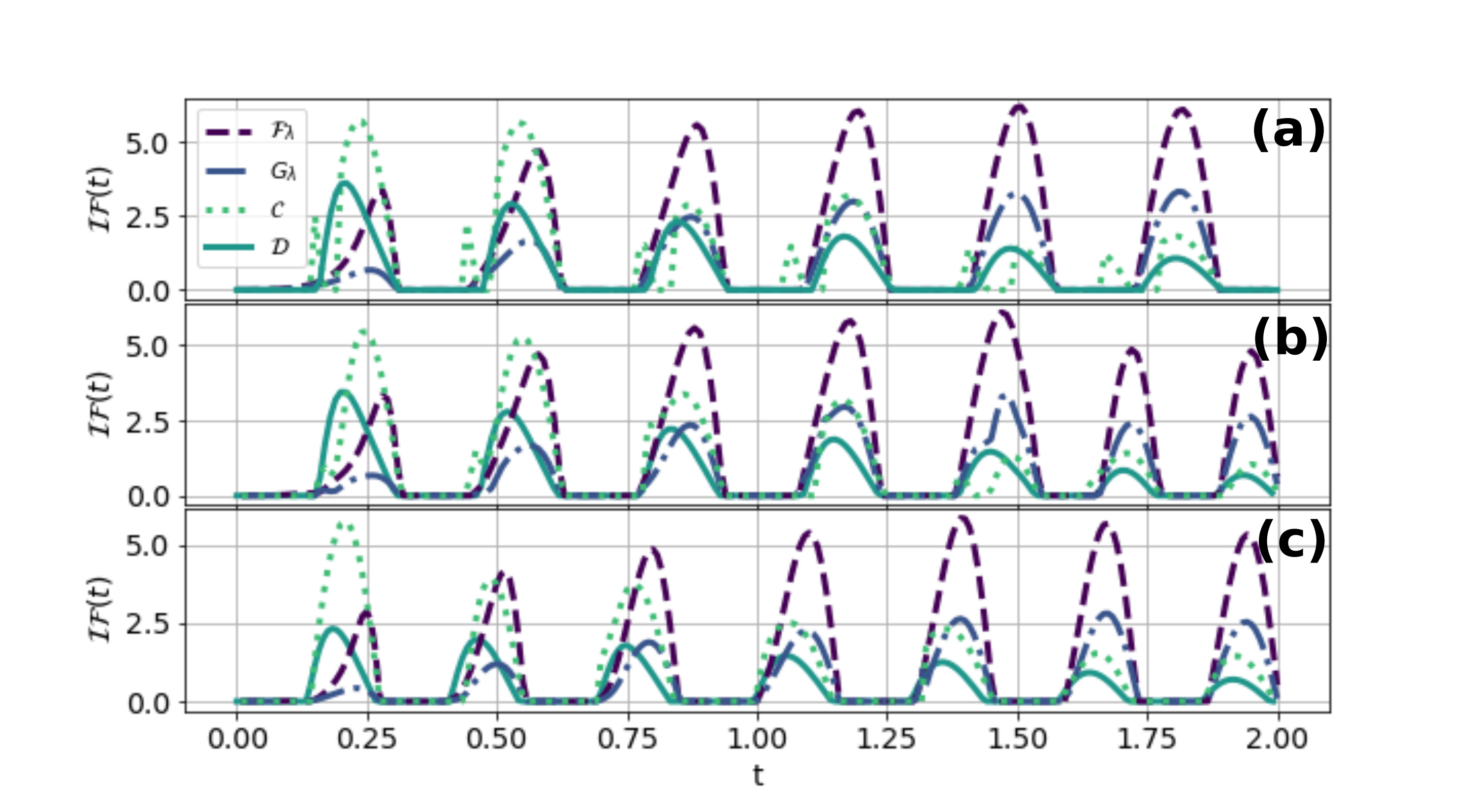}
\begin{footnotesize}
\caption{Incoming flows for $\mathcal{F}_{\lambda}(t)$, $G_{\lambda}(t)$, $\mathcal{D}(t)$ and $\mathcal{C}(t)$ as a function of time for the following set of different situations \textbf{(a)}: No control. \textbf{(b)}: With control maximizing $\mathcal{F}_{\lambda}(T)$. \textbf{(c)}: With control maximizing $\mathcal{C}(T)$. Interaction parameters are fixed as $a_1=0.4$,  $a_2=0.6$, $\mathcal{R}=10$ and $T=2$, while $\lambda$ is set equal to 1 and the initial state is maximally entangled ($s=0$ and $\phi=0$). The POVM used for computing $G_{\lambda}$ from Eq. (\ref{povm}) was $P_{x}=\left \{E_1,E_2,E_3 \right \}$, where $E_1=\sqrt{2}/(1+\sqrt{2})\ket{10}\bra{10}$, $E_2=\sqrt{2}/(1+\sqrt{2})\ket{01}\bra{01}$ and $E_3=\mathbb{1} - E_1 - E_2$. Analogous results were obtained by fixing other interaction parameters.}
\label{optim}
\end{footnotesize}
\end{center}
\end{figure*}

An interesting point to notice is based on the fact that the concept of information flow may be applied to an entire set of different physical quantities, for instance, QFI, NM but also entanglement. So a natural question can be formulated: how are these information flows related to each other? In order to address such a question, we will use the control method presented before as a way to dive through the subspace of the best solutions and try to extract unknown relations from them. For simplicity, we will just focus our attention on the incoming flows, i.e. the time intervals where these quantities are an increasing function of time and analyze whether they are related or not within different controlled situations. With this purpose, we define the incoming flow $\mathcal{IF}(t)$ as

\begin{equation}
\mathcal{IF}(t)=\frac{dA(t)}{dt} > 0,
\end{equation}
where $A(t)=\big\{ \mathcal{F}_{\tilde{\lambda}}(t),G_{\tilde{\lambda}}(t),\mathcal{D}(t),\mathcal{C}(t) \big\}$, being $\mathcal{F}_{\tilde{\lambda}}(t)$ the QFI for some specific parameter $\tilde{\lambda}$, $G_{\tilde{\lambda}}(t)$ the Fisher information for a given POVM estimating the same specific parameter $\tilde{\lambda}$, $\mathcal{D}(t)$ the distinguishability between the two initial states that maximize BLP measure on Eq. (\ref{BLP}) and finally $\mathcal{C}(t)$ the concurrence between both atoms. Let us remark that all the above quantities are time dependent and is there dynamics what we intend to relate. As an illustrative and representative example, in Fig. \ref{optim} we plot all these incoming flows as a function of time for the case of environment estimation ($\tilde{\lambda}=\lambda$), and under the following set of different situations: without any control field (panel \textbf{(a)}), with a control field that maximizes $\mathcal{F}_{\lambda}(T)$ (panel \textbf{(b)}) and finally for a control protocol that maximally preserves $\mathcal{C}(T)$ (panel \textbf{(c)}), where T is a fixed evolution time.         

\renewcommand{\figurename}{Figure} 

Notably, something we can note from all the panels in Fig. \ref{optim} is based on the fact that there is an univocal dynamical relation between $\mathcal{F}_{\lambda}(t)$ for estimating a parameter of the environment and the revivals of distinguishability $\mathcal{D}(t)$, which is a common feature of NM and normally interpreted as a backflow of information that flows from the environment to the reduced open system. In this way, one could intuitively think that initially there is no information about the unknown parameter of the spectral density since to extract information about it, an interaction with the degrees of freedom of the environment must occur. But surprisingly, this interaction seems to be not a sufficient requirement unless we are in a time interval in which we are having a backflow of information, something which may only occur in a non-Markovian quantum evolution. As it is clear from all the panels in Fig. \ref{optim}, the only intervals in which we are gaining information about the environment is during the intervals where a backflow of information is occurring. Whenever the backflow stops and therefore we start to lose information, $\mathcal{F}_{\lambda}(t)$ decreases. 

\begin{figure}[!htb]
\begin{center}
\includegraphics[width=88mm,scale=0.99]{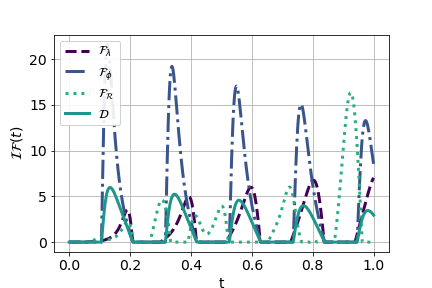}
\begin{footnotesize}
\caption{Incoming flows for $\mathcal{F}_{\mathcal{R}}(t)$, $\mathcal{F}_{\phi}(t)$, $\mathcal{F}_{\lambda}(t)$, and $\mathcal{D}(t)$ as a function of time for the non-controlled situation. Interaction parameters are fixed as $a_1=0.25$,  $a_2=0.75$, $\mathcal{R}=15$ and $T=1$, while $\lambda$ is set equal to 1 and the initial state is maximally entangled ($s=0$ and $\phi=0$). Same results were obtained by fixing other interaction parameters and initial states.}
\label{if}
\end{footnotesize}
\end{center}
\end{figure}

In a similar way, as it is shown in Fig. \ref{if}, the same dynamical relation between the QFI and NM arises when the estimation is for a parameter that characterizes the probe itself, such as its initial phase $\phi$. In this case, as the system starts to interact with the environment the information about the initial phase is leaked into it and is only due to the revivals of distinguishability and the backflow of information that the QFI of this initial phase increases back again. By the contrary, it is not surprising that on a situation where the estimation is for a parameter characterizing the interaction (such as $\mathcal{R}$), no backflow of information is needed and thus the QFI can increase independently of being on a non-Markovian time interval, as can be appreciated from the green dotted line in Fig. \ref{if}. Here the probe just needs to \textit{interact} (somehow) with the environment such as to extract information about the \textit{interaction}, without the necessity of receiving any backflow of information.

In summary, regarding QFI has a clear practical meaning and that there is no discussion with respect whether it is useful or not for specific tasks (i.e. for quantum metrology), this result linking QFI and NM gives the last a clear and indisputable meaning as a quantum resource, either for estimating a parameter of the environment or even to maximally preserve the information of a parameter characterizing the reduced open system \cite{bib:bylicka2013non,bib:env_resource2,bib:anand2019quantifying,bib:berk2019resource,bib:mirkin_ent,bib:mirkin2019information} \footnote{A minor comment but also important to highlight is that considering the QFI is defined from a maximization over all possibles measurements, it is crucial to show that there exists at least one particular POVM that exhibits the same dynamical behaviour. This is shown for $G_\lambda$ in all panels of Fig. \ref{optim}.}. 

\renewcommand{\figurename}{Figure} 

Let us go back now to Fig. \ref{optim} and focus on the controlled scenarios that are being studied. First, in the case where we pursue the maximization of the functional $\mathcal{F}_{\lambda}(T)$ in panel \textbf{(b)}, it can be seen that the information flows corresponding both to QFI and $\mathcal{D}(T)$ are accommodated by the control in order to exhibit a maximum value for $\mathcal{F}_{\lambda}(T)$ at the end of the protocol, as it has already been discussed in the previous subsection.

Finally, in regards to entanglement and its interplay between QFI and NM, in the scenarios covered by panel \textbf{(a)} and \textbf{(b)} in Fig. \ref{optim}, it is clear that there is no relation at all between these three quantities. For instance, what panel \textbf{(a)} is telling us is that the concurrence $\mathcal{C}(t)$ between the atoms may increase independently of being in a time interval in which a backflow of information is being manifested. This is not surprising since it is well known that a common Markovian environment may induce some degree of non-controlled entanglement between two non-interacting parts coupled to it \cite{bib:mirkin2019information,bib:ent_markov1,bib:ent_markov2}. With respect to panel \textbf{(b)}, where the control task is to optimize the final value of $\mathcal{F}_{\lambda}(T)$, as it has been previously shown in Figs. \ref{met1} and \ref{optim_ov}, entanglement is not the crucial factor for enhancing the parameter estimation. So it is not surprising to not have a clear correlation in this particular situation. However, the most interesting point to stress arises from panel \textbf{(c)}, where the control task consists now on the preservation of entanglement at time T. In this case, all the quantities are clearly correlated. The best way that the control field finds to preserve entanglement is by taking advantage of whenever a backflow of information is occurring, such as to recover from this backflow the entanglement that was previously washed out when the information was being lost into the environment.

\section{Final remarks}
In this work, we have studied the problem of quantum metrology under the framework of open quantum systems within a non-Markovian quantum evolution. The main motivation was to deepen into the relationship between apparently disconnected physical quantities, such as QFI, QSL, NM and entanglement. By addressing this complicated problem under a particular but fully analytical and controlled physical system, we have first shown that the speed of evolution and the accuracy of estimation are deeply connected. In this way, by exploring the process of estimation for a whole variety of different classes of parameters, the best initial metrological state proved to be the one that evolved fastest and not the maximally entangled one. This was shown by the use of the QSL.

Another important key-point of our work was based on the possibility of controlling externally the composite quantum system used as a probe. Under this context, by maximizing the value for the QFI at a given final evolution time, we have shown how the total QFI flow throughout the evolution can be exploited and accommodated by the control field in order to achieve the best precision of estimation at that final evolution time, which may be of experimental interest. This could be certainly useful in recent experiments \cite{bib:xu2019readout}. 

At last, by using the optimal control method not as an end in itself, but as a tool to explore the best solutions and extract unnoticed relations from them \cite{bib:mirkin_ent,bib:mirkin2019information}, we have focused on the dynamics of QFI, NM and entanglement to see whether these quantities were correlated or not under different control schemes. In all the scenarios considered, we have found a direct dynamical relation between the incoming flows of QFI (both for environment and phase estimation) and the revivals of distinguishability, which gives NM a concrete use as a resource for quantum metrology. In regards to entanglement and its dynamical interplay with QFI and NM, we have shown that when we optimize a control field to maximally preserve the initial entanglement at a fixed final evolution time, the incoming flows of entanglement coincide perfectly with the incoming flows of QFI and distinguishability. In other words, as  information is being recovered from the environment, via the revivals of QFI and distinguishability, this backflow is here used by the control to retrieve the entanglement that was previously washed out when the information was being leaked into it.

With the results on our back, we sincerely expect this work to shed light and clarity on the problem of quantum metrology and its sharp connection with QSL, NM and entanglement.  

\begin{acknowledgements}
The work was partially supported by CONICET (PIP 112201 50100493CO), UBACyT (20020130100406BA), ANPCyT (PICT-2016-1056), and National Science Foundation (Grant No. PHY-1630114).
\end{acknowledgements}

\bibliography{main.bib}

\end{document}